\documentclass[journal]{IEEEtran}
\usepackage{cite}
\usepackage{authblk}
\usepackage{graphicx}
\usepackage{caption}
\usepackage{amsmath}
\usepackage{tabularx}
\usepackage[linesnumbered,ruled,vlined]{algorithm2e}
\usepackage{algpseudocode}
\usepackage{xcolor}
\usepackage{amsfonts}
\usepackage[bottom]{footmisc}
\usepackage{subcaption}
\usepackage{comment}
\usepackage{multirow}
\usepackage{lscape}
\usepackage{pifont}
\include{pythonlisting}
\usepackage{cite}

\ifCLASSINFOpdf

\else

\fi

\begin{document}
\title{Differentially Private Enhanced Permissioned Blockchain for Private Data Sharing in Industrial IoT*}   
\author[1]{Muhammad Islam}
\author[2]{Mubashir Husain Rehmani}
\author[3]{Jinjun Chen}
\affil[1, 3]{Swinburne University of Technology, Hawthorn, VIC 3122, Australia}
\affil[2]{Munster Technological University, Rossa Avenue, Bishopstown, Cork, Ireland}
\renewcommand\Affilfont{\itshape\small}
\maketitle
\thispagestyle{empty} 
\pagestyle{empty} 
\begin{abstract}
The integration of permissioned blockchain such as Hyperledger fabric (HF) and Industrial internet of Things (IIoT) has opened new opportunities for interdependent supply chain partners to improve their performance through data sharing and coordination. The multichannel mechanism, private data collection and querying mechanism of HF enable private data sharing, transparency, traceability, and verification across the supply chain. However, the existing querying mechanism of HF needs further improvement for statistical data sharing because the query is evaluated on the original data recorded on the ledger. As a result, it gives rise to privacy issues such as leaking of business secrets, tracking of resources and assets, and disclose of personal information. Therefore, we solve this problem by proposing a differentially private enhanced permissioned blockchain for private data sharing in the context of supply chain in IIoT which is known as (EDH-IIoT). First, we integrate differential privacy into the chaincode (smart contract) of HF which evaluates the query and adds a calibrated noise into it. Second, we propose an algorithm to efficiently utilize $\epsilon$ through reuse of the privacy budget for the repeated queries. Third, we also propose an algorithm to track the privacy budget ($\epsilon$) and avoid the degrade of privacy preservation in case of multiple queries on the same portion of the ledger's data. Furthermore, the reuse and tracking of $\epsilon$ enables the data owner to ensure that $\epsilon$ does not exceed the threshold which is the maximum privacy budget ($\epsilon_{t}$). Finally, we model two privacy attacks namely linking attack and composition attack to evaluate and compare privacy preservation, and the efficiency of reuse of $\epsilon$ with the default chaincode of HF and traditional differential privacy model, respectively. The results confirm that EDH-IIoT obtains an accuracy of 97\% in the shared data for $\epsilon$ = 1, and a reduction of 35.96\% in spending of $\epsilon$.
\end{abstract}
\vspace{0.2cm}
\IEEEpeerreviewmaketitle
\renewcommand{\thefootnote}{\fnsymbol{footnote}} 
\footnotetext[1]{A preliminary version of this paper has been accepted in $12^{th}$ EAI International Conference on Broadband Communications, Networks, and Systems, 2021.} 
\section{Introduction}
\label{sec:intro}
\begin{figure}[bhp]
\begin{center}
\includegraphics[width=\linewidth]{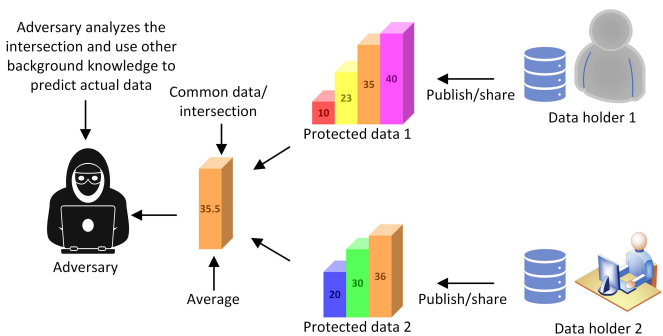}
\caption{Demonstration of composition privacy attack.}
\label{fig:ca}
\end{center}
\end{figure}

Industrial Internet of Things (IIoT) is the subclass of IoT in the industrial sector which enables various devices, sensors, and controllers to collect and process data \cite{iiot}. Similarly, blockchain is another emerging technology which was introduced by Satoshi Nakamoto in 2008 for virtual currency \cite{papforsatoshi}. Blockchain is distributed ledger in which each transaction is recorded, and each peer duplicates the ledger in a consistent manner. The consistency is achieved through a consensus mechanism in which every transaction is cryptographically verified and added sequentially to the chain which is agreed by all participants of the network \cite{bcforiiotreview}. In this way, it maintains a distributed, consistent, and temper-proof record with the attractive features of transparency, verification, and traceability. Generally, based on the participants access to the shared ledger, blockchain can be classified into two types which are permissioned and permissionless. In permissioned blockchain, the shared ledger can be accessed by selected participants whereas in permissionless blockchain each participant can access the shared ledger and take part in consensus \cite{whenwhichbc}\cite{appofbciniot}. \\
The integration of blockchain and IIoT has opened new opportunities for supply chain model in IIoT to share data and coordinate activities. The interdependent partners in supply chain such as supplier manufacturer, distributor, and retailer share their resources and information to improve individual performance, and achieve common goals aligned with their individual aims and objectives \cite{bcimprovsupply}.  Among the two types of blockchain, permissionless blockchain has got many issues including scalability, privacy, and low transaction processing speed which is why it is not a good option for IIoT \cite{appofbciniot}. On the other hand, permissioned blockchain such as HF is an attractive option for supply chain scenario in IIoT because of high transaction processing rate, known participants, scalability, and confidential transactions \cite{hyper}. Therefore, a lot of works including but not limited to \cite{datashrframforIIoT,privsharing,blockwithappstore,materialtrackingcat1} have adopted Hyperledger fabric for private data sharing between the consortium of parties/organizations. In supply chain scenario, multichannel and private data collection along with transient field of transaction enable confidential information sharing between groups of similar interest \cite{hyper}. However, the focus of the mentioned works is on confidential data sharing, i.e., the data is confined between the sharing parties. The utilization and analysis of data recorded on the ledger have not been discussed. Therefore, it makes it not suitable for supply chain scenario in which the interdependent parties share their individual or group data with others along with the guarantee that their sensitive information will be protected.\\
A channel in HF limits the communication between the valid group members which belong to that channel. Furthermore, for each such channel, a separate ledger (private ledger) is maintained. For example, in supply chain, manufacturer-distributer and distributor-retailer come together to make two separate groups. The channels between the groups limit the data sharing and exchange of confidential information between the valid members. For instance, in \cite{privsharing}, private channel (multichannel) and data collection mechanisms of HF have been used to enable private communication between groups in the context of smart cities. Similarly, the transient field of transaction is used to hide private data from other participants which are not members of the channel. In addition, the querying mechanism of HF enables applications and partners/organizations in the chain to send queries against the private ledgers of individual partners. The query response is generated from the data (transactions) recorded on the ledger. However, the data sharing through querying mechanism gives rise to privacy issues including leak of business secrets, tracking of resources and assets, and personal information of individuals.\\
For example, two different distributors (possible competitors) in the chain can distribute products to retailers in a region. To track the distribution and manage the demand response in real-time, the distributor needs the statistics regarding the product including number of items sold, and peak hours visits of customers. The distributor sends a numerical query to the concerned retailer for total number of items sold. This query is evaluated on the data recorded on the retailer ledger and sent back to the distributor. Two types of privacy attacks are possible namely linking attack and composition attack. In linking attack, the information learned is combined with the background knowledge to reveal actual data \cite{linkingattack}. Similarly, in composition attack, the adversary combines the results published by two different data holders, i.e., two supply chain participants and perform matching to reveal actual data \cite{compattackdef1}\cite{compattackdef2} as shown in Fig. \ref{fig:ca}. From the numerical answer, the distributor can launch linking attack to get information regarding transactions with another distributor. Moreover, the transactions can also be used for learning the lifestyle, spending trends and financial status of individual in the region. \\ 
Similarly, in another scenario, the responses of the same query/published results from two supply chain participants linked through the same channel in HF are evaluated on the ledger copy (channel members have identical copies). The adversary can launch a composition attack by sending same queries to both participants and perform matching or intersection on the results which gives rise to privacy issues, i.e., the responses can be combined to reveal the true answer. Therefore, the existing querying mechanism in HF needs further improvement. The privacy leak in statistical queries is extensively studied by using differential privacy. Differential privacy is a well-known privacy preserving technique which was introduced by C. Dwork for statistical databases \cite{dpint2006, dwork2006, dwork2014algorithmic, tran2019privacy}. The basic idea of differential privacy is that the removing or adding of a single data record from a dataset does not significantly change the output of an algorithm (applied to the dataset) under the constraint of differential privacy. \\
To achieve differential privacy, a calibrated noise is added to the actual data. The amount of noise is controlled by the parameter $\epsilon$ which is known as differential privacy budget. Furthermore, the adversary model is stronger than other privacy preserving techniques i.e., the adversary has all other information except his target person/data record. However, in case of multiple queries on the same dataset, the privacy protection degrades due to accumulation of privacy budget which results in high value of $\epsilon$ \cite{trackbudget}. Similarly, the adversary can send repeated queries to the same dataset and take the average to predict the actual data \cite{dwork2008differential}. Moreover, the number of queries which the data holder can answer is bounded by $\epsilon$. Consequently, this problem becomes more prominent in blockchain-based supply chain in the context of IIoT. The reason is that the same ledger data (dataset) is queried by multiple participants, i.e., manufacturer as well as distributor send queries to purchase history database of retailer for statistical analysis. Furthermore, in practical scenarios, the same data is required by multiple organizations/companies connected in the chain. As a result, it increases the risk of revealing the sensitive information. Therefore, the traditional differential privacy model needs improvement in terms of efficient utilization of privacy budget $\epsilon$ in order to increase the number of queries the data holder can answer with the guaranteed privacy. Furthermore, it also needs improvement for the scenario of repeated queries \cite{trackbudget}.\\
Consequently, private data sharing in the scenario of supply chain using HF has three main challenges which are as following: first, the privacy issues in query mechanism of HF due to both linking and composition attacks; second, degrade of privacy preservation due to repeated queries on the same dataset (ledger data in proposed case); third, inefficient spending of the privacy budget $\epsilon$. To leverage the benefits of blockchain, enable private data sharing and analysis of data recorded on the ledger in the scenario of supply chain in IIoT, this work comes with a solution to the above stated challenges. we propose a differentially private enhanced permissioned blockchain for private data sharing in the scenario of supply chain in Industrial IoT which is known as EDH-IIoT. In EDH-IIoT, we integrate differential privacy into the chaincode (smart contract) of HF. An initial version of this work is available in \cite{islam2021differential}. However, it did not consider the privacy risk due to composition attack, the degrade of privacy protection due to repeated queries, and inefficient spending of privacy budget $\epsilon$. Hence, in this work, we address a new problem of degrade of privacy protection due to multiple/repeated queries on the same dataset (ledger data). In addition, we propose an algorithm to categorize the queries into new and repeated queries. For repeated queries, the proposed algorithm reuses the privacy budget which improves the spending of maximum or threshold differential privacy budget $\epsilon_{t}$ and enables the data holder to answer more queries. Furthermore, we model two privacy attacks namely linking attack and composition attack to investigate the privacy preservation of the proposed EDH-IIoT. The contributions of this work are as following:
\begin{itemize}
\item We propose a differentially private enhanced permissioned blockchain for private data sharing in the scenario of supply chain in IIoT which is known as EDH-IIoT. Through EDH-IIoT, we enable transparency, traceability, verification, and analysis of the data recorded on the blockchain ledger across the supply chain. Furthermore, to overcome the inefficiency in the spending of privacy budget $\epsilon$ and avoid the degrade of privacy preservation through repeated queries on the same dataset (ledger's data), we propose two algorithms.
\item We amalgamate differential privacy into the chaincode (smart contract) of HF which first accesses the ledger's data and then a calibrated noise is added to the actual data before sharing it with the requester which guarantees privacy preservation through $\epsilon$-differential privacy. Furthermore, we model two privacy attacks namely linking and composition attack to investigate the privacy preservation of the proposed EDH-IIoT.
\item We enable reuse of the privacy budget $\epsilon$ through algorithm 1 for the repeated queries on the same dataset (ledger data), whereas algorithm 2 is used to ensure that the accumulated $\epsilon$ does not exceed the threshold $\epsilon_{t}$ which is the maximum privacy budget. Moreover, through evaluation, we show that the EDH-IIoT obtains 97\% of accuracy in the shared data and 35.96\% reduction in spending of the privacy budget $\epsilon$.
\end{itemize}    
The remaining part of the paper is structured as following: Section \ref{sec:letr} presents a detailed literature review of the related works. Similarly, Section \ref{sec:pw} presents the proposed work (EDH-IIoT) in detail. Furthermore, evaluation and simulation results are discussed in Section \ref{sec:eval}. Finally, conclusion of the proposed work is presented in Section \ref{sec:con}.  
\section{Literature Review}
\label{sec:letr}
In this section, we divide the literature review into three categories which are private data sharing using HF, privacy preserving data sharing in IIoT, and efficient utilization of differential privacy budget. In the following, we discuss each category of work in detail. 
\subsection{Private Data Sharing Using HF}
\label{subsec:pptHLF}
In \cite{datashrframforIIoT}, a secure data sharing framework for Industrial data has been proposed. To increase the efficiency of blockchain-based data sharing, data has been divided into private data, community data and public data. Furthermore, private data is encrypted, and it has the highest level of privacy following by community data whereas public data can be shared in open public. Moreover, the proposed framework consists of three layers which are data layer, detection layer and blockchain layer. The data collected through data layer is classified by using community detection algorithm of detection layer based on the scope of data sharing. Finally, the blockchain layer using HF records the data. However, the query is evaluated through chaincode on ledger data using the default query mechanism of HF which gives rise to privacy issues as discussed in Section \ref{sec:intro}. Similarly, in \cite{privsharing}, HF based private data sharing framework for smart city has been proposed. The proposed framework divides smart city into many communicating parties using multichannel model and private data collection of HF. Therefore, it restricts the communication and access of data to valid members of the channel. Each channel has distinct types of data such as health and medical data, smart car, financial and smart energy. Furthermore, privacy and confidentiality are achieved through data private data collection i.e., encryption and hashing.
Furthermore, in \cite{blockwithappstore}, a reputation mechanism for the consortium of app stores have been proposed which is called TrustChain. HF has been used as a scalable solution along with privacy preservation of app store users. Furthermore, cloud is used to collect reputation events and send it to the blockchain network for recording. The app threat level and reputation of users are identified by processing the events by accessing it from blockchain network. Finally, the scalability of the proposed system is evaluated to validate that the system performs as intended. Apart from this, in \cite{materialtrackingcat1}, a material tracking mechanism in the context of manufacturing in IIoT has been proposed. In the proposed mechanism, IIoT and HF have been combined to automate the activities and enable tracking and verification across supply chain. Furthermore, the privacy of sensitive and confidential data is enabled through distributed ledger in which each participant has a copy of the validated transactions.  Moreover, selective data sharing is enabled in the semi-trusted environment of supply chain i.e., each participant shared portion of its data with others while retain other information which is confidential. \\
In \cite{bprpds}, authors have proposed a blockchain-based privacy-preserving and rewarding private data-sharing scheme (BPRPDS) for IoT. The proposed scheme comprehensively addressed privacy issues in selling data and getting rewards. However, the scheme relies on security features of cryptographic algorithms and unlikability of transactions. Similarly, due to the high processing time of such algorithms, the latency of transactions also increases. As a result, it is not suitable for a real-time data sharing environment where the users expect minimum response time.    

\begin{figure*}[htp]
\begin{center}
\includegraphics[scale=0.4]{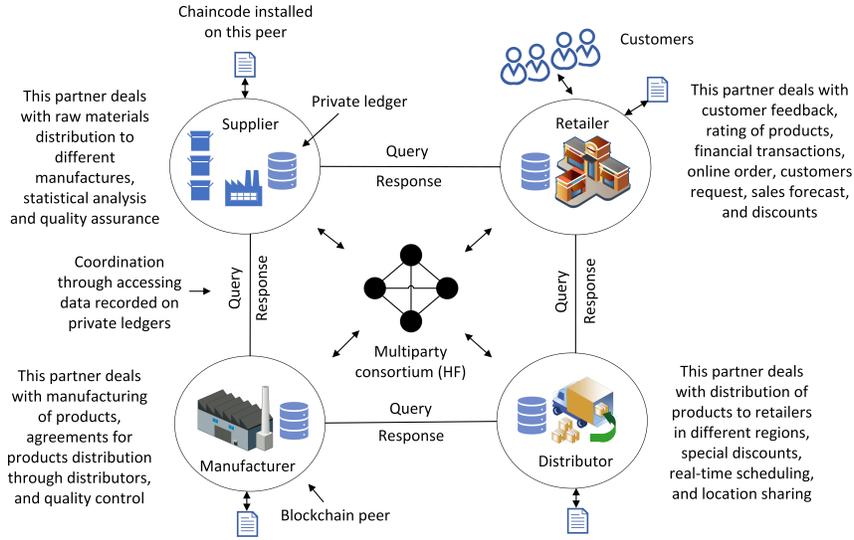}
\caption{Private data sharing and accessing scenario in EDH-IIoT.}
\label{fig:scenario}
\end{center}
\end{figure*}

\subsection{Privacy Preserving Data Sharing in IIoT}
\label{subsec:ppdsIIoT}
The work in \cite{bcandfedforIIoT} has proposed a blockchain-based secure architecture for data sharing in the scenario of multiple parties in IIoT. The main idea of this work is that a local model is trained on the perturbed data and then shared with other parties/nodes rather than raw data. Similarly, a global model is trained in a federated environment whereas the individual private data is not revealed to others. Permissioned blockchain is used for controlling and managing the access to the data. Similarly, a data sharing mechanism between multiple parties in IIoT has been proposed in \cite{ppdatashareinIIoT}. In contrast to the aforementioned works, this work does not consider blockchain. A three-party model is proposed which consists of data owner/workers, service provider and data consumer. The service provider collects data from workers which is perturbed before sharing with the data consumer in order to protect the sensitive information of individuals. Two scenarios were proposed based on the relation between service provider and data consumer. In the first scenario, service provider and consumer form a coalition model in which both parties share identical information. In the second scenario, both acts as independent parties and service provider perturb the data before sharing it with the data consumer.\\
In another work \cite{reputationsytemforconsret}, a reputation management scheme has been proposed for retailer-consumer channel in the scenario of supply chain in IIoT. The consumer anonymously gives feedback for the purchased products which is further used for reputation management of retailers. The proposed scheme uses randomized signature and zero-knowledge proof techniques to achieve anonymization of individual reviews. Furthermore, the architecture is built on top of blockchain which provides transparency in the reputation management. Similarly, the work in \cite{reputationscheme} has proposed an incentive mechanism for maintaining positive reputation among the mining nodes in the consensus phase of blockchain. Through the proposed reputation mechanism, the nodes in IIoT are categorized into normal and abnormal based on their cooperation behavior in the network. Furthermore, normal nodes are encouraged to take part in consensus whereas abnormal nodes are punished to decrease the malicious activities in the system. However, the mentioned works have focused on privacy preservation and confidential exchange of information. The utilization and analysis of data recorded on the ledger data has not been investigated.\\
In \cite{eppda}, an efficient privacy-preserving data aggregation mechanism for federated learning (EPPDA) has been proposed. The proposed mechanism has effectively addressed the reverse attack on model training. However, EPPDA cannot protect against privacy attacks other than the reverse attack on model training. Therefore, it does not fit well in our proposed scenario. Similarly, authors in \cite{pp_socialiot} has proposed a new hybrid privacy-preserving method for federated learning scenarios in social IoT systems. The proposed method combines encryption and Bayesian differential privacy to avoid private information leakage on data-level and content-level. However, the method relies on centralized entity to aggregate the results which is vulnerable to privacy attacks. In \cite{encsmargrid}, an encryption based trusted execution environment is proposed for data sharing in smart grid. However, in the proposed threat model, data owner is required to trust the data utility (data user). Also, it does not apply to query-based data analysis because the data is encrypted. Therefore, it is not suitable for the scenario considered in this work. 
\subsection{Efficient Utilization of Differential Privacy Budget}
\label{subsec:dpbtr}
Differential privacy has been adopted as an optimistic technique to enable privacy preserving data sharing. Numerous woks have adopted differential privacy included but not limited to \cite{dwork2016concentrated,bun2016concentrated,dwork2015generalization,mcsherry2007mechanism,abadi2016deep}. One of the attracting features of this technique is the consideration of strongest adversary model, i.e., the adversary is assumed to know all other information except its target person or actual information which makes it preferable as compared to other privacy preserving techniques. However, often the same dataset is queried multiple times in practical scenarios. In this case, the privacy budget $\epsilon$ accumulates for all such queries \cite{differentialpubsurvey}. As a result, the privacy protection degrades with increasing number of queries.\\
To overcome this problem, the data owner records the spending of privacy budget to ensure that the accumulated privacy budget does not exceed the maximum limit or threshold. In literature, blockchain has been adopted to keep temper-proof records of data which is traceable and verifiable at the same time. In \cite{trackbudget}, a mechanism to track and control the spending of privacy budget has been proposed. Furthermore, to efficiently utilize the privacy budget, a mechanism of reusing the privacy budget has been adopted. The queries are categorized and recorded on the blockchain which is then used to decide whether to add fresh noise or reuse the previous noise. In this way, for repeated queries, the previous noise is utilized which avoids the excessive utilization of privacy budget. However, the proposed architecture combines blockchain and client-server model for data sharing which increases the overhead and latency. The HTTP requests are used to communicate with the server through smart contracts. Furthermore, the smart contracts execute on each peer which increases the risk of revealing the spending of privacy budget. \\
In \cite{semanawarepriv}, a semantic-aware privacy-preserving online location trajectory sharing mechanism (SEITP) has been proposed. The proposed mechanism effectively addressed the location privacy along with semantic privacy preservation. Although the proposed mechanism is suitable for online trajectory scenario, it does not address query-based data sharing, and privacy degradation with repeated queries.

\begin{table}[pb]
\caption{Key notations}
\begin{center}
\begin{tabular}{ll}
\hline
Notation & Explanation \\
HF & Hyperledger Fabric \\
$\epsilon$ & Differential privacy budget \\
$\epsilon_{t}$ & Maximum or threshold differential privacy budget \\
\textit{$T_{q}$} & Query transactions set with $N$ number of queries \\
\textit{$R_{q}$} & Response of $T_{q}$ with $N$ number of query answers \\
$\mu$ & Mean for Laplace distribution \\
$\lambda$ & Laplace scale \\
$\bigtriangleup$f & Sensitivity \\
$\epsilon_{rem}$ & Remaining $\epsilon_{t}$ \\
$\mathbb{G}$ & The queries recorded on the ledger \\
$N$ & Total number of queries \\
$l$ & Number of queries for which $\epsilon$ is reused \\
$z$ & Number of queries for which new $\epsilon$ is utilized \\
$\mathbb{C}$ & Number of items purchased in a single transaction \\
$r$ & Relative error in perturbed data \\
$a$ & Actual value of query response \\
$a^{'}$ & Perturbed value of query response \\
$L_{d}$ & Private ledger of the blockchain peer \\ 
\hline
\end{tabular}
\label{tab:notation}
\end{center}
\end{table}

\section{Differentially Private Enhanced Permissioned Blockchain for Private Data Sharing in the Context of Supply Chain in IIoT (EDH-IIoT)}
\label{sec:pw}
In this section, the proposed work is presented in detail. A data sharing and accessing scenario in the scenario of supply chain in IIoT based on HF is shown in Fig. \ref{fig:scenario}. Furthermore, this section also discusses the system model and threat model followed by the privacy preservation and working of the proposed EDH-IIoT. Moreover, in this Section, the supply chain partner which sends request for data sharing to other partners in the chain is called data requester whereas the supply chain partner which responds to the data sharing request is called data provider. Table \ref{tab:notation} summarizes other key notations used in this Section.

\begin{figure}[htp]
\begin{center}
\includegraphics[width=\linewidth]{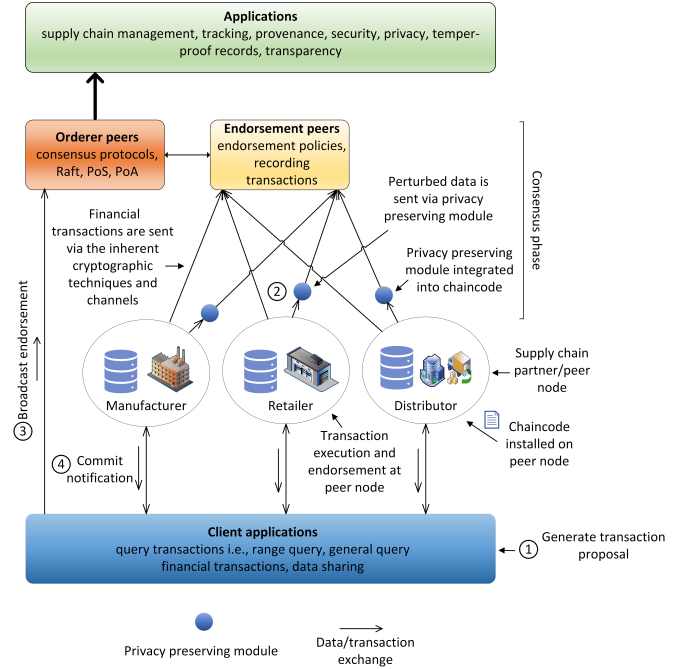}
\caption{Transaction flow and relationship between various components in EDH-IIoT.}
\label{fig:sm}
\end{center}
\end{figure}

\subsection{System Model}
\label{subsec:sm}
Our proposed system model is depicted in Fig. \ref{fig:sm}. A differential privacy-based privacy preserving module is proposed and integrated into the chaindcode (smart contract) of HF. In this way, it is installed on each peer. Furthermore, the system model consists of client applications, endorsing peers, privacy preserving module, orderer peers, and the resultant applications on top of these components. The details of each functioning module and transaction types are given below.
\subsubsection{Client applications}
Client application is the component in the proposed model which sends transactions to the blockchain peers or blockchain network. The client application uses the services provided by peers i.e., write to the ledger, search the ledger, query the ledger, update the ledger, and read from the ledger. Supply chain partners use this component to access data from other partners/peers. More specifically, it invokes the chaincode installed on each peer to perform the aforementioned operations or services. As a result, all participants of the blockchain network use this component for the mentioned operations.
\subsubsection{Endorsing peers}  
The endorsing peers are a subset of blockchain peers in the network that performs endorsement of the ledger proposals/updates. The proposed ledger update or chaincode invocation requested by client applications is approved by the endorsing peers.
\subsubsection{Ordering peers}
The orderer peers have two responsibilities which are collection of the transactions from all peers in the network and carry out consensus. HF provides a range of options for consensus mechanisms such as Solo, Kafka and Raft. The transactions are added to the blocks which are then sent for validation where it is further appended to the chain. 
\subsubsection{Privacy preserving module}
Privacy preserving module is the component which implements differential privacy through Laplace mechanism and adds calibrated noise to the query results. We propose two chaincode (smart contract) functions to handle query transactions. In the first function, we implement our proposed algorithm \ref{alg:algreuse} whereas the second function implements the proposed algorithm \ref{alg:alg1}. Furthermore, it also handles the reusing, spending, and tracking of privacy budget $\epsilon$. The working of privacy preserving module is given in Section \ref{subsec:ppt}.
\subsubsection{Transactions}
In the proposed system model, the transactions are of two types which are write transactions and query transactions. Write transactions further include financial transactions, and update transactions. Similarly, query transactions include querying or reading transactions, and range transactions.\\
Furthermore, supply chain participants (partners) use client applications to share and access data from other partners in the network. Therefore, it enables a coordination scenario through data sharing. Moreover, the financial transactions are shared by using the default features of blockchain such as encryption. However, the query transactions are handled in a different way, i.e., noise is added to the query response. In addition, for query transactions, we only consider statistical COUNT and SUM queries in this work. For example, how many customers have got more than 10\% discounts on the purchase of a particular product? and what is the total number of purchased items? We leave the consideration of other query types to the future work. The query transactions are denoted as $T_{q}=\{f_{1}, f_{2}, f_{3}..f_{N}\}$ in which $f_{1}, f_{2}, f_{3}..f_{N}$ denote the queries to be evaluated on the ledger data. Moreover, $N$ is the total number of query transactions whereas each query transaction consists of one query. Similarly, the noise added responses to the quires are denoted as $R_{q}= \{f^{*}_{1}, f^{*}_{2}, f^{*}_{3}… f^{*}_{N}\}$. 
\subsection{Threat Model}
\label{subsec:tht}
In our threat model, the adversary is a competitor (i.e., more than one distributor) of another partner in the same supply chain. As a result, the competitor is interested to know the business secrets of its counter partner with others. The proposed coordination environment favors the adversary in the sense that it can send queries to one of the private group members which includes the competitor. As a result, the queries (numerical queries in this work) are evaluated on the private ledger of that group. From the answers, the adversary can learn regarding the business secrets i.e., special discounts, and contracts. Two types of attacks are considered which are linking attack and composition attack. In the first scenario, the adversary launches linking attack to get the intended knowledge. In the linking attack, the learned knowledge (observed data) is combined with the background knowledge (from other sources) to reveal actual information (i.e., prices, discounts, and contracts). Similarly, in the second scenario, the adversary launches composition attack in which data sharing request is sent to members of the same channel in HF. As a result, the queries results evaluated on the same data (channel members have same copy of the ledger). The adversary tries to find matching records by performing intersection of data shared by two members. Consequently, from the intersection, the adversary tries to reveal actual data (i.e., target person).\\
Furthermore, we consider a strong adversary i.e., it has all other knowledge except its target i.e., actual information. The competitor is honest-but-curious which means that it cannot edit or delete the data but can share it with others. In another scenario, the linking attack can also be used for revealing the lifestyle, spending trends, and financial status of the individuals in the region. The reason is that the queries are evaluated on the ledger which consists of transactions including the local residents in the region. Moreover, this learned information can be provided to third parties such as advertising companies for financial benefit. 
\subsection{Differential Privacy}
\label{dp}
Differential privacy is a popular and widely adopted privacy preservation technique in the context of statistical analysis. The core idea of differential privacy is that the change caused by the presence or absence of a single record in a dataset to the output of an algorithm (applied to the dataset) is bounded by a constant $e^{\epsilon}$ under the constraint of differential privacy. According to \cite{dpint2006}, the definition of differential privacy is as following:\\\\
\textbf{Definition 1:} \textit{A randomized function or algorithm Q satisfies $\epsilon$-differential privacy if for all datasets $D_{x}$, $D_{y}$ which differs in one record, and for all $S \subseteq Range(Q)$, the following holds \cite{dpint2006}:}
\begin{align}
P[Q(D_{x}) \in S] \leq e^{\epsilon} \times P[(Q(D_{y})\in S] && \text{(by \cite{dpint2006})}
\label{eq:eqdp}
\end{align}
Where Range(Q) denotes the set of all possible outputs of function Q. $\epsilon$ is known as differential privacy budget, and S is a subset of Range(Q). Furthermore, a smaller value of $\epsilon$ gives good privacy and vice-versa.\\
To achieve differential privacy, a calibrated noise is added to the actual data. Moreover, the calibrated noise is generated through two well-known mechanisms which are exponential and Laplace \cite{differentialpubsurvey}. However, we use the Laplace mechanism which is best fit for numerical queries. According to \cite{lapdwork}, the distribution function of Laplace mechanism is given as following:
\begin{align}
f(x, \mu, \lambda) = \frac{1}{2\lambda}e^{\frac{-|x-\mu|}{\lambda}} && \text{(by \cite{lapdwork})}
\label{eq:lappr}
\end{align} 
Here, $\lambda$ is known as Laplace scale and $\mu$ is the mean for the Laplace distribution. Furthermore, $\lambda = \frac{{\bigtriangleup}f}{\epsilon}$ in which $\bigtriangleup$f is known as \textit{sensitivity} and defined as the maximum difference of the results of two queries on adjacent datasets $D_{x}$ and $D_{y}$. For $D_{x}$ and $D_{y}$ differing in one record it is denoted as following \cite{dpint2006,dwork2014algorithmic}:
\begin{align}
{\bigtriangleup}f = {| f(D_{x})-f(D_{y}) |}_{1} && \text{(by \cite{dpint2006, dwork2014algorithmic})}
\label{eq:sensitivity}
\end{align} 
Furthermore, often several queries are sent to be evaluated on the target dataset (ledger data in the proposed work). In that case, each query is not guaranteed according to $\epsilon$-differential privacy. The reason is that with each query, the value of privacy budget $\epsilon$ increases and consequently, the privacy guarantee decreases. According to \cite{differentialpubsurvey}, this property of differential privacy is known as sequential composition which is given in the following theorem.\\\\
\textbf{Theorem 1:} \textit{If a set of $m$ algorithms  $\textbf{Q} = \{Q_{1}, Q_{2}, Q_{3}..Q_{m}\}$ is applied to the same dataset, and each algorithm $Q_{i}$ satisfies $\epsilon$-differential privacy then $\textbf{Q}$ satisfies ($m.\epsilon$)-differential privacy \cite{differentialpubsurvey}.}\\\\
Consequently, care must be taken regarding spending of privacy budget $\epsilon$ to get ensure that it provides the minimum required guarantee of privacy preservation. We explore this topic further in the context of proposed work in the next Section.

\begin{figure}[tp]
\begin{center}
\includegraphics[width=\linewidth]{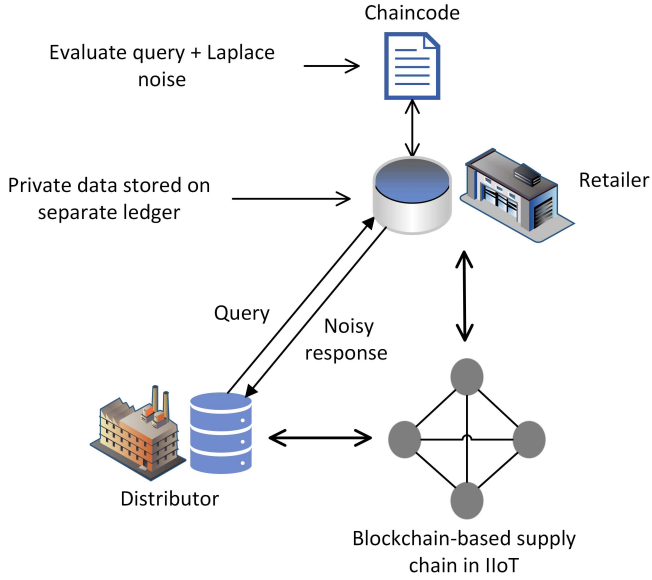}
\caption{Illustration of query evaluation and noise addition through chaincode on private data in EDH-IIoT.}
\label{fig:sc}
\end{center}
\end{figure}

\subsection{Working of Differentially Private Enhanced Permissioned Blockchain for Private Data Sharing in the Context of Supply Chain in IIoT (EDH-IIoT)}
\label{subsec:wmech}
In this section, we will use the term peer for a node which represents an organization or partner of the supply chain in IIoT. Furthermore, an organization can have more than one peers. Similarly, the peers record data of an organization which is accessed by peers of other organization through queries (query transactions). In HF, the collection of transactions from different peers, sequencing, and addition of transactions into blocks are the responsibilities of orderer peers. Similarly, the endorsing and validating peers perform endorsement and validation, respectively. This concept makes the HF architecture modular and enables it to validate transactions faster and hence achieve higher throughput. To this end, working of the proposed EDH-IIoT is categorized into four steps which are (1) transaction proposal, (2) transaction endorsement and privacy preservation, (3) execution of ordering service, and (4) transaction validation and commit. The sequence of the four steps in the system model is shown in Fig. \ref{fig:sm}. In the following, each of these steps is discussed in detail. 
\subsubsection{Transaction Proposal and Endorsement}
In the first step, application client invokes the chaincode installed on each peer by sending a transaction proposal. The endorsement policy of the chaincode specifies the peers which are called endorsing peers i.e., a single peer on the channel can perform endorsement or all members of the channel are required to endorse a transaction proposal. The transaction proposal serves as an input for the chaincode. In response, if the proposed chaincode invocation is valid then the chaincode generates transaction proposal response which is then shared with the client application. Furthermore, in this step, the ordering service is not executed, and the interaction is limited to only client applications and endorsing peers. Moreover, the state of the ledger is not changed in this step because the ordering peers are not involved.
\subsubsection{Execution of Privacy Preserving Module}
In the second step (after invocation of chaincode in the first step), privacy preserving module is executed as demonstrated in Fig. \ref{fig:sm}. Each peer executes its instance of the chaincode to evaluate the query response using the private ledger records (data). Similarly, algorithms \ref{alg:algreuse} and \ref{alg:alg1} are executed in the given sequence to reuse privacy budget $\epsilon$, and generate the calibrated noise for each query $f_{i}$ which is added to the actual query result of the evaluation. We leave the detailed working steps of algorithms \ref{alg:algreuse} and \ref{alg:alg1} to the next Section. The demonstration of query evaluation followed by the noise addition using the chaincode is shown in Fig. \ref{fig:sc}. Finally, the perturbed query response $f_{i}^{*}$ is sent back to the client application. Other transactions such as financial transactions and ledger update transactions are not handled through privacy preserving module of the chaincode. Therefore, transactions other than statistical query transactions follow the same procedure of the default transaction flow of HF.
\subsubsection{Execution of Ordering Services}
In step 3, the endorsed transaction proposals are coupled with the transaction and sent to the ordering peers by application clients. The ordering peers collect transactions from other client applications in the network and perform consensus (Solo, Kafka and Raft). The consensus mechanisms are not the focus of this work, so we leave its consideration to the future work. Therefore, we proposed a Solo consensus mechanism in which only a single ordering peer performs all the responsibilities of the ordering service. Finally, the consensus phase results in packaging of transactions into blocks which are then distributed to all peers for validation. 
\subsubsection{Transaction Validation and Commit}
In the final step, the orderer peers broadcast blocks to the validating peers. Each validating peer verifies every transaction carried by the block to confirm that it meets the endorsement policy. Similarly, after passing the validation phase with success, the blocks are committed and appended to the chain on each peer. Furthermore, the invalid blocks (which fail the validation phase) are retained for audit and not appended to the chain. Finally, notifications of the successful processing of the last two steps are then sent to the corresponding client applications. 

\begin{algorithm}
\small
\let\oldnl\nl
\newcommand{\nonl}{\renewcommand{\nl}{\let\nl\oldnl}}
\caption{Proposed algorithm for reusing the differential privacy budget $\epsilon$ in EDH-IIoT}
\label{alg:algreuse}
\SetAlgoLined
\DontPrintSemicolon
\nonl
\textbf{Input:} Ledger data $L_{d}$, $\mathbb{G}$ (queries already answered and recorded on the ledger), query transactions with $N$ queries $T_{q} = \{f_{1}, f_{2}, f_{3}…f_{N}\}$\;
\nonl
\textbf{Output:} Differential private (perturbed) query responses $R_{q} = \{f^{*}_{1}, f^{*}_{2}, f^{*}_{3}… f^{*}_{l}\}$, where $l$ is the number of queries out of $N$ for which the privacy budget is reused and $l \leq N$\;
\nonl
\textbf{Initialization:} Iteration \textit{i = 1}\;
\While {\textit{Invoke}} {
	\While {$i \leq l$}{
		Get the attributes of query from the input query transaction\;
		Search the query $f_{i}$ in $\mathbb{G}$\;
		\If (\tcp*[h]{successful match}){$f_{i} \in \mathbb{G}$} { 
  			\textit{$f_{i} \leftarrow f^{*}_{i}$}\tcp*[h]{Table \ref{tab:catofquery}}\;
		}\Else {
			\KwRet \tcp*[h]{the query is not found}
		} 
		\textit{$i \leftarrow i + 1$}\;
}

	\KwRet $R_{q} = \{f^{*}_{1}, f^{*}_{2}, f^{*}_{3}… f^{*}_{l}\}$\;

} 
\end{algorithm}

\begin{table}[htp]
\caption{Query categorization on the basis of attributes. Cat denotes the category.}
\begin{center}
\begin{tabular}{|c|c|c|c|c|}
\hline
\textbf{Attributes} & \textbf{$Cat_{1}$} & \textbf{$Cat_{2}$} & \textbf{$Cat_{3}$} & \textbf{$Cat_{4}$} \\
\hline
Privacy budget used & $\epsilon_{1}$ & $\epsilon_{2}$ & $\epsilon_{3}$ & $\epsilon_{4}$ \\
\hline
Query response & $f_{1} \leftarrow f^{*}_{1}$ & $f_{2} \leftarrow f^{*}_{2}$ & $f_{3} \leftarrow f^{*}_{3}$ & $f_{4} \leftarrow f^{*}_{4}$ \\
\hline
\end{tabular}
\label{tab:catofquery}
\end{center}
\end{table}

\subsection{Privacy Preservation, Reusing and Tracking of the Privacy Budget $\epsilon$}
\label{subsec:ppt} 
In this section, the mechanism of privacy preservation, reusing, and tracking of differential privacy budget $\epsilon$ are discussed. In Section \ref{subsec:tht}, we presented the details that how a competitor in the same supply chain can act as an adversary and tries to steal or reveal business secrets and personal information of customers. To avoid the risk of being exposed, we adopted differential privacy which is integrated into the chaincode. Therefore, the queries sent by client application (possible adversary) are evaluated using the privacy preserving model instead of the bare evaluation function of the chaincode (default setting). The superiority of differential privacy is that the small amount of noise addition does not change the output by a significant amount in statistical queries (COUNT, SUM, AVERAGE) as evident from the equation \ref{eq:eqdp}. On the other hand, the receivers of the query response including adversary get the perturbed response instead of the actual response. Therefore, the data requester with normal behavior (non-adversarial behavior) gets the intended knowledge from the query responses. However, the adversary intention to get the knowledge of his interest is avoided. \\
Similarly, another advantage of differential privacy is that the adversary is assumed to have strong background knowledge i.e., the worst scenario of privacy disclosure. Consequently, it maintains the beneficial knowledge in the shared data whereas the privacy preservation of individual records is guaranteed. For example, if the actual response of the query which asks for the number of customers who got more than 10\% discounts on the purchase of a particular product is 500 then a noise of magnitude 2 will make it 502. It is evident that the number 502 is still beneficial for the data requester because it gives almost the same knowledge. However, for an adversary, the number 502 is a confusing response because he is not sure whether the target person is included or excluded? \\ 
In this way, privacy leakage through linking attack is avoided. Similarly, in the scenario of composition attack, the attacker sends queries to participants of the same channel of HF. For example, the distributor (distributor A) acts as an adversary and sends queries to the retailer and another distributor (distributor B) in the same geographical region, i.e., postcode, city etc. The retailer is further sharing channel with distributor B. Therefore, the data shared between channel members is recorded on respective ledgers of both participants. Distributor A sends the same queries to distributor B and the results from both participants are analyzed for the results of common queries. For example, the number of products purchased by a customer or group of customers will appear in both the results. Although, it will be slightly different due to random noise addition by both participants. However, the adversary can infer the correct answer, i.e., by taking an average of both results. The proposed EDH-IIoT avoids the risk of exposure through composition attack by searching the query in the recorded queries through algorithm \ref{alg:algreuse}. If the query is answered before then it will be recorded on the ledger and hence listed on the ledgers of both participants, i.e., retailer, and distributor B. In this way, new noise addition is avoided, and the same query answer is returned to distributor A. Distributor A will not be able to find matching queries through intersection. Consequently, the risk of exposure through composition attack is avoided. The detailed discussion and working of algorithm \ref{alg:algreuse} is presented in Section \ref{subsub:reuse}.\\  
However, along with these advantages, the increase of privacy budget $\epsilon$ with increase in the number of queries evaluated on the same dataset (ledger data in this work) is a disadvantage of differential privacy as shown by \textit{Theorem 1} in Section \ref{dp}. Several queries on the same dataset (ledger data in the proposed case) give rise to excessive utilization of privacy budget $\epsilon$, i.e., the privacy spending is accumulated which results in a higher value of $\epsilon$. As a result, the privacy preservation of sensitive data degrades. Furthermore, the data provider/owner needs to ensure that the accumulated differential privacy budget does not exceed the maximum privacy budget $\epsilon$. In EDH-IIoT, we reuse the privacy budget to overcome the inefficiency and excessive spending of $\epsilon$. In the following, we describe the mechanism of reusing the differential privacy budget $\epsilon$ in EDH-IIoT.

\begin{algorithm*}
\small
\let\oldnl\nl
\newcommand{\nonl}{\renewcommand{\nl}{\let\nl\oldnl}}
\caption{Proposed algorithm for adding Laplace noise and tracking of privacy budget $\epsilon$ in EDH-IIoT}
\label{alg:alg1}
\SetAlgoLined
\DontPrintSemicolon
\nonl
\textbf{Input:} Ledger data $L_{d}$, query transactions with $N$ number of queries $T_{q} = \{f_{1}, f_{2}, f_{3}…f_{N}\}$\;
\nonl
\textbf{Output:} Differential private query responses $R_{q} = \{f^{*}_{1}, f^{*}_{2}, f^{*}_{3}… f^{*}_{z}\}$, where $z$ is the number of queries out of $N$ for which fresh Laplace noise is added to generate the perturbed response\;
\nonl
\textbf{Initialization:} Iteration \textit{j = 1}, threshold privacy budget $\epsilon_{t}$, remaining privacy budget $\epsilon_{rem} = \epsilon_{t}$, \textit{x} is a random variable, \textit{noise} = 0, mean $\mu$ = 0, sensitivity $\bigtriangleup$f = 100, Laplace scale $\lambda = \frac{{\bigtriangleup}f}{\epsilon}$\;
\While {\textit{Invoke}} {
     Classify the query $f_{j}$ as $f_{j}$ with equal/unequal distribution of $\epsilon$ on the bases of requester's  		 privacy preservation requirement\;
     Identify the privacy requirements of each data requester \tcp*[h]{known participants in permissioned   		 		 blockchain} \;
     Make a strategy for each data requester by assigning values to $\epsilon_{f_{j}}$ according to equation  				 \ref{eq:epforeq} and \ref{eq:split}\;	
		\While {$j \leq z$}{
		\Indp 
			\If (\tcp*[h]{tracking of $\epsilon$ according to expressions \ref{ex:accbudget} and \ref{ex:enq2}}){$						\epsilon_{rem} > 0$} {   
				\textbf{Call} QueryFunction($\epsilon_{f_{j}}$)\;
				$\epsilon_{rem} \leftarrow \epsilon_{rem} - \epsilon_{f_{j}}$\;
				Write $\epsilon_{rem}$, $\epsilon_{f_{j}}$, and query transaction to the ledger \tcp*[h]{according to 					Table \ref{tab:catofquery}}\; 
		    }
			\Else {
				Stop executing query $f_{j}$\;
				\KwRet\; 
			}
			\textit{$j \leftarrow j + 1$}\;
	   }
	\nonl
	$\textbf{FUNCTION}\rightarrow QueryFunction(\textit{$\epsilon_{f_{j}}$})$\;  
	\Indp 
		Evaluate query $f_{j}$ on the original ledger data $L_{d}$\;
		\textbf{Call} LaplacianFunction($\epsilon_{f_{j}}$)\;
		Add noise to perturb the query response $f^{*}_{j} \leftarrow f_{j} + noise$\;
		\KwRet {$f^{*}_{j}$}\;
		\Indm
		\nonl
		$\textbf{FUNCTION}\rightarrow LaplacianFunction(\textit{$\epsilon_{f_{j}}$})$\;  
		\Indp 
			Generate Laplacian noise using \textit{$f(x;\mu,\frac{{\bigtriangleup}f}{\textit{$\epsilon_{j}$}})$} 				through equation \ref{eq:lappr}\; 
			\KwRet {$noise$}\;
		\Indm
	\KwRet $R_{q} = \{f^{*}_{1}, f^{*}_{2}, f^{*}_{3}… f^{*}_{z}\}$\;
} 
\end{algorithm*}

\subsubsection{Reuse of Differential Privacy Budget $\epsilon$ in EDH-IIoT}
\label{subsub:reuse}
We adopt the concept of \cite{trackbudget} for reusing and tracking of privacy budget $\epsilon$ with the necessary modification in the architecture. In \cite{trackbudget}, the client-server model has been adopted to share data in which the smart contract is used to decide whether to utilize fresh privacy budget or send the previous answer to the requester. In contrast, our dataset consists of the ledger data and the smart contract accesses the data of the local ledger as shown in Fig. \ref{fig:sc}. We denote the total number of queries as $N$ and $i^{th}$ query as $f_{i}$. Furthermore, $l$ is the number of queries out of $N$ for which the privacy budget is reused where $l \leq N$, i.e., if all queries sent by requesters are found as repeated queries, then $l = N$, otherwise, $l \leq N$. Consequently, out of $N$ queries, $N-l$ queries need fresh Laplace noise for perturbation. Similarly, the list of queries which has already executed and recorded on the ledger is denoted as $\mathbb{G}$. Each query $f_{i}$ in $\mathbb{G}$ is recorded with the associated budget $\epsilon_{i}$ spent in generating its perturbed response. According to our system model in Fig. \ref{fig:sm}, each transaction which invokes the chaincode function carries a query $f_{i}$. Furthermore, the queries are categorized according to the attributes. For example, the color of the product in query transactions is used to categorize the queries. As a result, color of the product represents a separate category of a statistical query as shown in Table \ref{tab:catofquery}. It is to be noted that this categorization is temporary, and the color attribute is adopted because the proposed context represents a supply chain where the products purchase transactions of different types are recorded on blockchain. Therefore, it can be changed accordingly, i.e., any other attribute of the product can be used as differentiating factor between queries. \\
The query transactions invoke the chaincode function which implements algorithm \ref{alg:algreuse}. The attributes of the query $f_{i}$ are extracted and a search in $\mathbb{G}$ is performed. On 100\% match of the attributes, the query response $f^{*}_{i}$ is returned to the requester. In this way, the previous privacy budget is utilized and spending of fresh privacy budget $\epsilon$ from the total or threshold budget $\epsilon_{t}$ is avoided. The step-by-step execution is shown in algorithm \ref{alg:algreuse}. The attributes extraction and search of query $f_{i}$ in $\mathbb{G}$ are performed in lines 2 and 3 of algorithm \ref{alg:algreuse}, respectively. Similarly, if a successful match of $f_{i}$ is found in $\mathbb{G}$ then the previous response $f^{*}_{i}$ is returned as shown in lines 4 to 6. Moreover, if the query $f_{i}$ is not found in $\mathbb{G}$ then the chaincode function returns and stops further execution which is performed in line 9 of algorithm \ref{alg:algreuse}. In this case, the query $f_{i}$ is sent for the first time and needs fresh Laplace noise to generate its perturbed repose $f^{*}_{i}$. In the proposed work, fresh noise addition is performed by the second privacy preserving function of the chaincode which implements algorithm \ref{alg:alg1}. We discuss the execution of algorithm \ref{alg:alg1} in the following Section.
\subsubsection{Tracking of Differential Privacy Budget $\epsilon$ in EDH-IIoT}
\label{subsub:tracking}
The second problem of spending of differential privacy budget is to ensure that the accumulated privacy budget $\epsilon$ does not exceed the threshold budget $\epsilon_{t}$. Therefore, we enable tracking of the spending of privacy budget to avoid the degrade of privacy preservation. In the proposed EDH-IIoT, the total privacy budget $\epsilon_{t}$ is the threshold set by each data provider in the supply chain. According to \textit{Theorem 1}, this budget splits among the total number of queries. Therefore, it results in two use cases which are (a) equal distribution of privacy budget, and (b) unequal distribution of privacy budget. 
\paragraph{Distribution of the threshold $\epsilon_{t}$ for equal privacy levels}
In this use case, the request for data sharing comes from different participants of the supply chain with the same privacy preservation level, i.e., the data provider treats all requests with equal level of privacy preservation. Therefore, all requesters which send requests for data sharing are equally trusted, and the data sharing provider does not consider difference between the queries from requesters with respect to the privacy preservation guarantee. As a result, $\epsilon_{t}$ is distributed equally among all the queries sent by the requesters. Consequently, for $N$ number of queries the distribution is as following:
\begin{align}
\epsilon_{f_{k}} &= \frac{\epsilon_{t}}{N} && \text{(by \textit{Theorem 1 \cite{differentialpubsurvey}})}
\label{eq:epforeq}  
\end{align} 
Where $\epsilon_{f_{k}}$ represents the privacy budget for $k^{th}$ query and $k \in \{1,2,3...N\}$.\\\\
As a result, the data provider accumulates the spending of privacy budget for each query and compares it with the threshold privacy budget $\epsilon_{t}$. More specifically, it ensures that the following inequality holds before a new query $f_{k+1}$ is answered.
\begin{align}
\sum_{k=1}^{N} \epsilon_{f_{k}} \leq \epsilon_{t}  && \text{(for given threshold $\epsilon_{t}$ \cite{trackbudget})}
\label{ex:accbudget}
\end{align}  
Where $\sum_{k=1}^{N} \epsilon_{f_{k}}$ is the sum of $N$ fractions of privacy budget $\epsilon_{f_{k}}$ spent on each query $f_{k}$.\\\\
Consequently, this technique enables the data provider to ensure that the privacy preservation is guaranteed as intended.
\paragraph{Distribution of the threshold $\epsilon_{t}$ for unequal privacy levels}
In this use case, a total of $N$ number of queries is sent from requesters. However, the data provider treats them with different privacy preservation requirements. For instance, if manufacturer sends $w_{1}$ whereas a distributor sends $w_{2}$ number of queries to a retailer, then $N = w_{1} + w_{2}$. Furthermore, the retailer wants to answer with different privacy preservation guarantees. For example, the retailer wants a higher privacy guarantee for distributor than manufacturer because of the adversarial history of distributor. In this case, each query $f_{k}$ is answered with a different fraction of privacy budget $\epsilon_{f_{k}}$ which depends on the profile of data requester, i.e., the adversarial history.  For $N$ number of queries by requesters, the accumulation of all fractions of privacy budget $\epsilon_{f_{k}}$ is given as following: 
\begin{align}
\epsilon_{sum} = \sum_{k=1}^{N} \epsilon_{f_{k}} && \text{(by \textit{Theorem 1 \cite{differentialpubsurvey}})}
\label{eq:split}
\end{align} 
Where $\epsilon_{sum}$ and $\epsilon_{f_{k}}$ represent the total consumption of privacy budget and fraction privacy budget spent on $k^{th}$ query $f_{k}$, respectively.\\\\
Similarly, the data provider ensures that the following inequality holds before a new query $f_{k+1}$ is answered.
\begin{align}
\epsilon_{sum} \leq \epsilon_{t} && \text{(for given threshold $\epsilon_{t}$ \cite{trackbudget})}
\label{ex:enq2}
\end{align}  
Where $\epsilon_{sum}$ is the sum of total privacy budget spent in answering $N$ number of queries.\\\\ 
Similarly, $z$ is the number of queries out of $N$ which need fresh Laplace noise for perturbation. Furthermore, $z \leq N$, i.e., if all queries from data requesters are different and sent for the first time then $z = N$, otherwise, $z \leq N$. Consequently, $N-z$ is the number of queries for which the privacy budget $\epsilon$ is reused. The step-by-step execution is shown in algorithm \ref{alg:alg1}. On invocation of the chaincode function, the $f_{j}$ is classified as query $f_{j}$ with equal/unequal distribution of privacy budget $\epsilon$ in line 2 of algorithm \ref{alg:alg1}. Furthermore, the data provider identifies the privacy preservation requirements and makes strategy for each data requester as shown in lines 3 and 4. The strategy is the agreement of both data sharing parties which may be set up prior to data sharing. Similarly, the remaining privacy budget $\epsilon_{rem}$ is checked to ensure that the data provider still has $\epsilon_{rem} > 0$. On success, the query is executed and the parameters $\epsilon_{rem}$, $\epsilon_{f_{j}}$, and the query transaction are written to the ledger as shown in lines 6-9 of algorithm \ref{alg:alg1}. Moreover, if the data provider has no budget to utilize then the query execution is stopped. Consequently, the data provider ensures that the privacy preservation is guaranteed as intended and that the utilized privacy budget $\epsilon_{sum}$ does not exceed the threshold $\epsilon_{t}$.  
\subsection{Complexity Analysis}
\label{sec:ca}
This Section discusses the time complexity of algorithms \ref{alg:algreuse} and \ref{alg:alg1} which have been defined as chaincode functions as described in Section \ref{subsec:sm}. The proposed algorithm \ref{alg:algreuse} consists of an outer \textit{While} loop which executes when the transactions set $T_{q}$ invokes the first chaincode function of privacy preserving module as discussed in Section \ref{subsec:sm}. Similarly, the transactions set $T_{q}$ consists of $N$ queries. The inner \textit{While} loop iterates according to $l$ number of queries where $l \leq N$. Furthermore, in the body of the inner \textit{While} loop, each statement takes \textit{O(1)}  time to complete. Consequently, in the worst case, all the $N$ queries will execute, i.e., all $N$ queries sent by the requesters are repeated queries. Accordingly, the time complexity of algorithm \ref{alg:algreuse} is \textit{O(N)}.\\
Similarly, on invocation of the second chaincode function of privacy preserving module, the outer \textit{While} loop of the algorithm \ref{alg:alg1} is executed. The inner \textit{While} loop then executes according to $z$ number of queries where $z \leq N$. Furthermore, in the body of the inner \textit{While} loop, each statement takes \textit{O(1)} time to complete. Hence, in the worst case, all $N$ number of queries will execute, i.e, all $N$ queries from requesters are different and sent for the first time. Therefore, the time complexity of algorithm \ref{alg:alg1} is \textit{O(N)}. It is evident from the time complexity analysis that the time increases linearly with the increase in the number of queries. Consequently, the high transaction processing rate of HF blockchain is maintained in our proposed EDH-IIoT.

\begin{table}[hbp]
\caption{Transaction body in the proposed EDH-IIoT}
\begin{subtable}[h]{0.48\columnwidth}
\centering
\caption{Write transaction}
\begin{tabular}{|m {6em} | m{6em} |}
\hline
contract id & contract version\\
\hline
contract function & time-out \\
\hline
\multicolumn{2}{|c|}{\textbf{contract arguments}}\\
\hline
product name & color \\
\hline
quantity & customer name \\
\hline
\end{tabular}
\label{tab:init}
\end{subtable}
\hfill
\begin{subtable}[h]{0.48\columnwidth}
\centering
\caption{Query transaction}
\begin{tabular}{|m{6em}|m{6em}|}
\hline
contract id & contract version\\
\hline
contract function & time-out \\
\hline
read only & - \\
\hline
\multicolumn{2}{|c|}{\textbf{contract arguments}}\\
\hline
customer name & - \\
\hline
\end{tabular}
\label{tab:query}
\end{subtable}
\label{tab:tranbody}
\end{table}

\begin{figure*}[tph]
\centering
\begin{subfigure}[t]{0.30\linewidth}
\centering
\includegraphics[width=\linewidth]{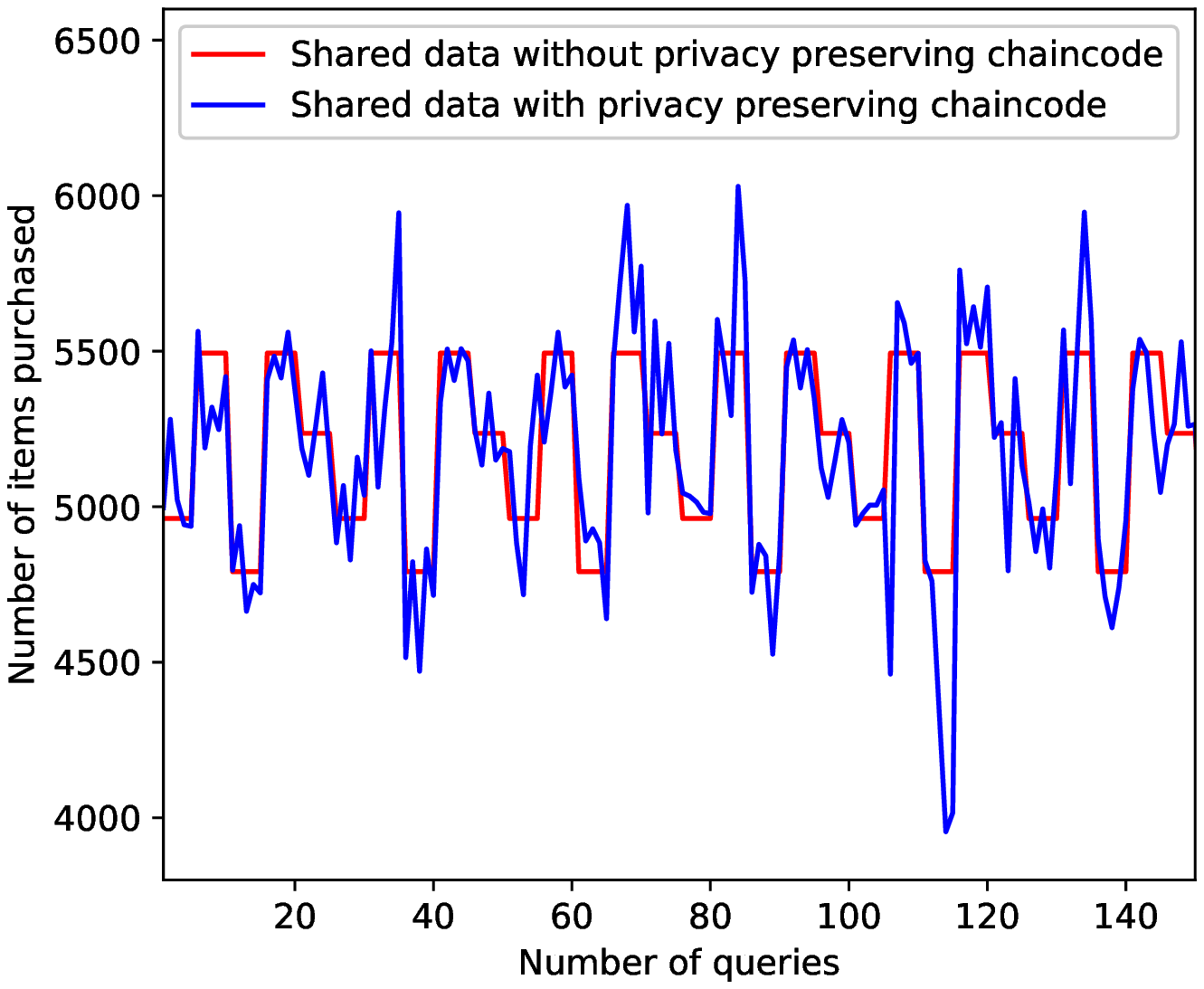}
\caption{Privacy budget $\epsilon_{t} = 1$}
\label{dpriv1}
\end{subfigure}
\begin{subfigure}[t]{0.30\linewidth}
\centering
\includegraphics[width=\linewidth]{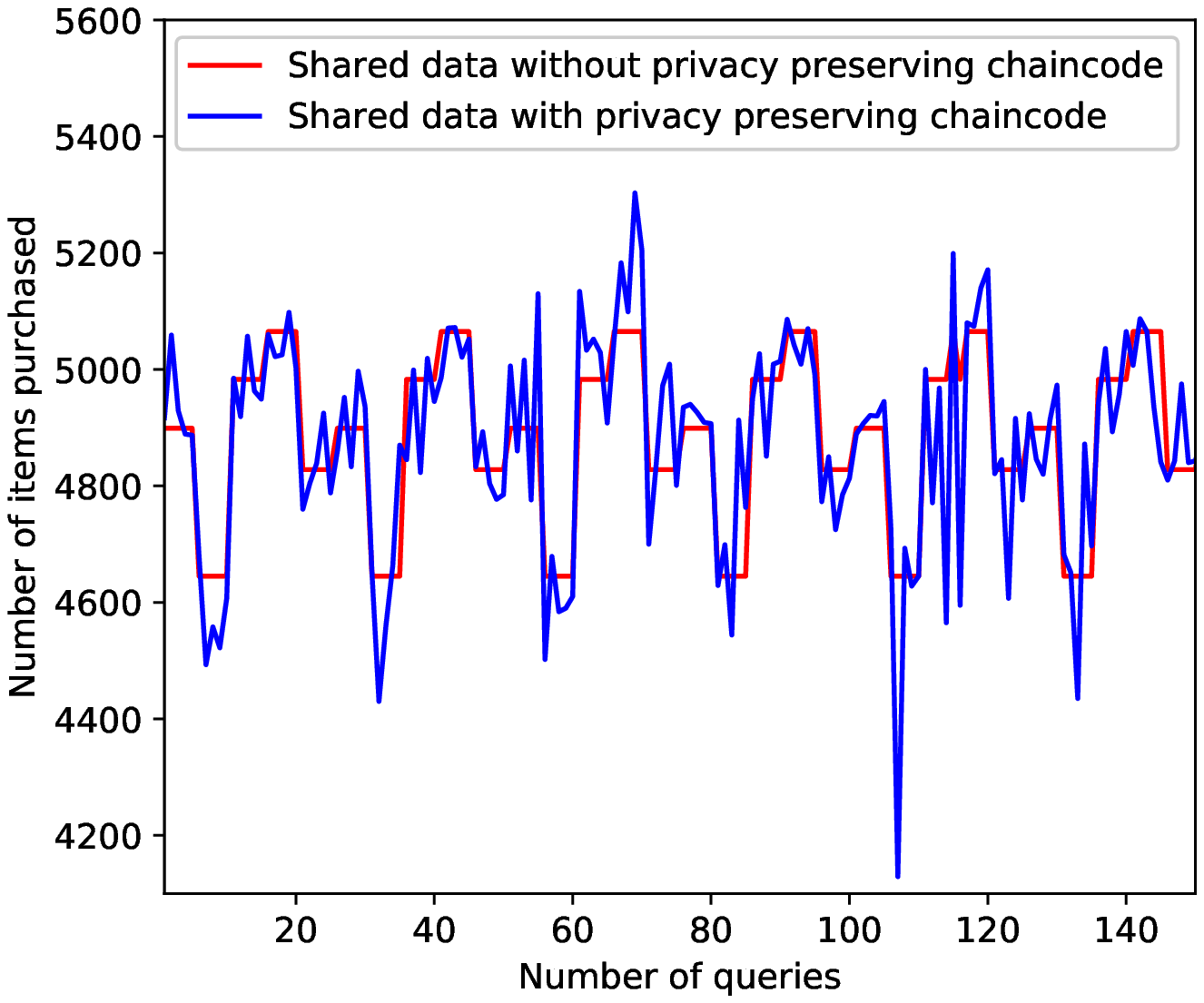}
\caption{Privacy budget $\epsilon_{t} = 2$}
\label{dpriv2}
\end{subfigure}
\centering
\begin{subfigure}[t]{0.30\linewidth}
\centering
\includegraphics[width=\linewidth]{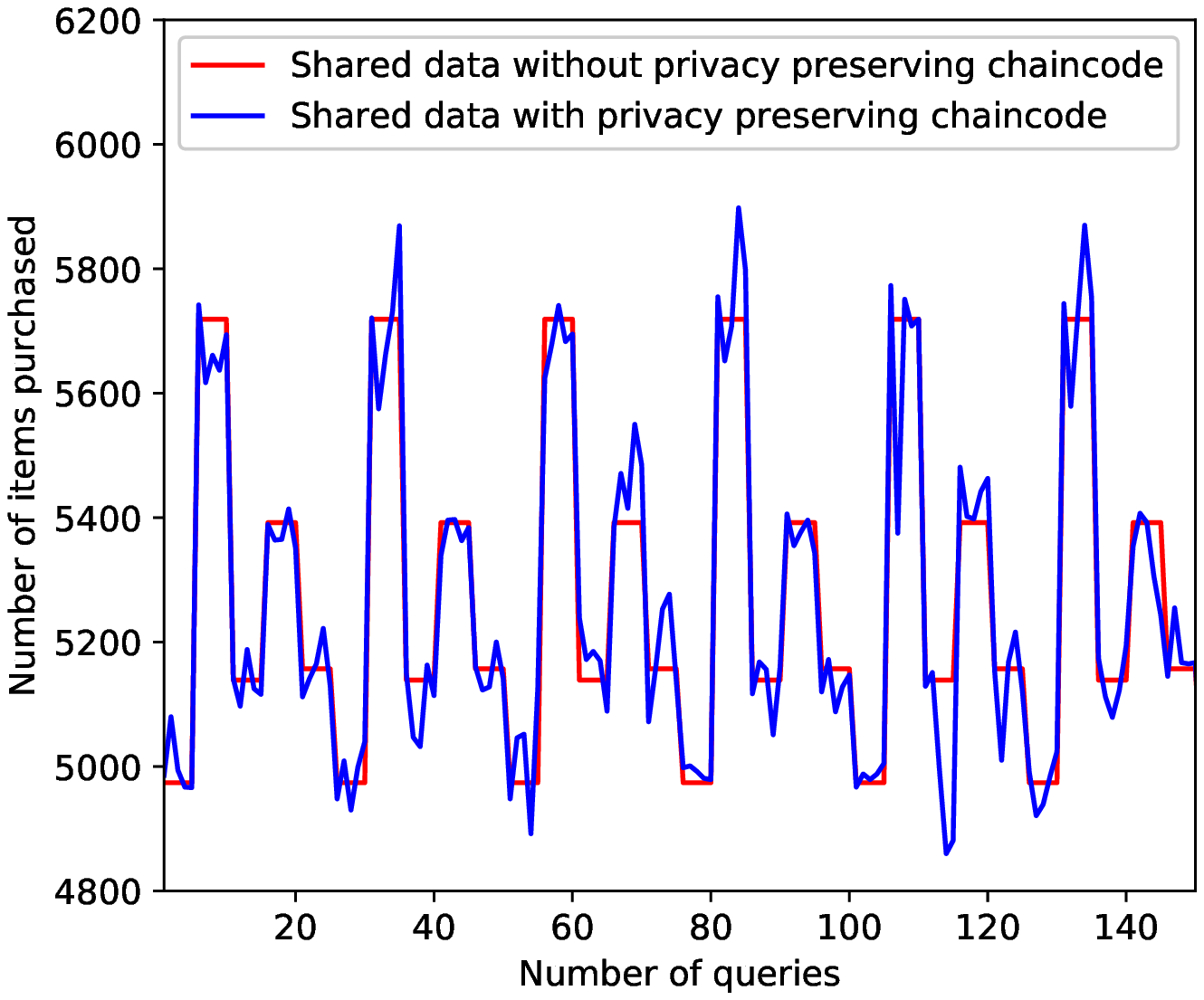}
\caption{Privacy budget $\epsilon_{t} = 3$}
\label{dpriv3}
\end{subfigure}
\caption{Comparison of privacy preserving chaincode (EDH-IIoT) with non-privacy preserving chaincode.}
\label{fig:priveval}
\end{figure*}

\section{Performance Evaluation}
\label{sec:eval} 
\subsection{Experimental Setup} 
\label{subsec:exs}
In this part of the paper, we present the performance evaluation of the proposed EDH-IIoT. The proposed blockchain network is simulated using HF which consists of two organizations each having a single peer. The organizations represent two supply chain partners which perform data sharing using query transactions as demonstrated in Fig. \ref{fig:sc}. The two-peer blockchain network is called software under test (SUT) which is evaluated for transaction processing in the proposed scenario. Furthermore, a single chaincode is installed on both peers in which a channel named as \textit{mychannel} is maintained between the peers. The endorsement policy in our proposed scenario requires endorsement of at least one member of the channel. Similarly, Caliper is used to test the SUT by sending the transactions \cite{caliper}. The SDK version of SUT is 1.4.11 whereas the Caliper version is 0.4.0. Furthermore, the proposed work is compared with the default settings of chaincode in HF. The software specifications include Ubuntu-18 64-bit operating system installed alongside Windows 10 whereas hardware specifications include a machine with Intel(R)Core (TM) i5-8250U CPU, 1.6 GHz of processor, and 8 GB of physical memory. \\
In the experiment, we simulated a two peer blockchain network repressing two organizations in the supply chain model. However, in real-world scenarios, a supply chain has many partners/organizations. As a result, it is interesting to scale the blockchain network to more than two organizations so that the overall effect can be evaluated. Therefore, we plan to evaluate a supply chain with more than two peers in our future work.
\subsection{Benchmark and Transaction Configuration}
\label{subsec:benchconfig}
The benchmark setting consists of two types of transactions which are write transactions and query transactions. Furthermore, The SUT initialization and querying are performed in two rounds. In the first round, the ledger is populated with 500 write transactions which consist of purchase transactions of five different customers given as \{Bob, Claire, David, Ali, Alice\}. In each transaction, $\mathbb{C}$ number of items of a particular product are purchased where $\mathbb{C} \in [1,100]$. Similarly, the transaction rate is varied according to 10-50 tran/sec. The write transactions are sent via a test with five workers. In the second round, query transactions are sent which are evaluated on the ledger state. Each transaction consists of a statistical query regarding the number of items purchased. Furthermore, query transactions are configured at a fixed rate of 10-50 tran/sec. The transactions configuration is given in Table \ref{tab:tranbody}. Moreover, the privacy budget threshold $\epsilon_{t}$ is varied between 1 to 5. In the proposed case, the transactions result in a dataset with rows and columns. Each row represents a single transaction whereas the columns consist of the fields given in Table \ref{tab:tranbody}. As a result, the maximum difference of the output of a SUM query (total number of purchased items) on two neighboring ledgers (datasets) differing in one transaction is 100. Therefore, according to \cite{trackbudget}, the sensitivity in equation \ref{eq:sensitivity} is given as $\bigtriangleup$f = 100. \\
In our benchmark setting, we only considered the protection of statistical data leakage. The reason is that our proposed EDH-IIoT is based on differential privacy which applies to statistical data. However, in real-world scenarios, sensitive data is a mix of both numerical and categorical data. Therefore, in the future, we plan to extend our work to include categorical data.
\subsection{Simulation Results and Discussion}
\label{subsec:simrsl}
The experiments are performed over five parameters which are (1) privacy preservation (2) relative error (3) tracking and spending of differential privacy budget $\epsilon_{t}$ (4) throughput, and (5) latency of transactions. In the following, we discuss the results obtained for each parameter.  
\subsubsection{Privacy Preservation}
\label{subsec:pp}
In the proposed EDH-IIoT, the supply chain data providers maintain two types of ledgers for data which are public and private. All members of the blockchain network see the data recorded on a public ledger. Similarly, each channel has a separate private ledger which is accessible to members of the channel. In the proposed scenario, distributor accesses private data recorded on the ledger of the retailer through query transactions as shown in Fig. \ref{fig:sc}. The ledger is initialized with transactions as described in Section \ref{subsec:benchconfig}. The query is regarding the total number of items purchased of a particular product. The retailer needs to share data with distributor as well as preserve its privacy, i.e., business secrets, and sensitive information of associated customers. Therefore, the query is evaluated by the chaincode on private ledger of the retailer and calibrated noise is added to the result by using algorithm \ref{alg:alg1}. The threshold privacy budget $\epsilon_{t}$ is varied according to $\epsilon_{t} \in \{1,2,3\}$.\\

\begin{figure}[pb]
\begin{center}
\includegraphics[width=\linewidth]{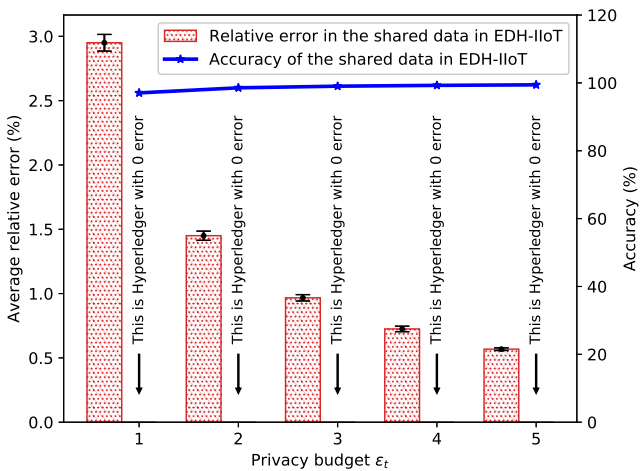} 
\caption{Relative error in query response with varying differential privacy parameter $\epsilon$. The results are within 95\% of confidence interval.}
\label{fig:relerr}
\end{center}
\end{figure}

A total of 150 queries are sent to the retailer's peer and the responses are generated using the default setting (non-privacy preserving) of the chaincode in HF and chaincode in EDH-IIoT. The results obtained are plotted for three different values of $\epsilon_{t}$ as shown in Fig. \ref{fig:priveval}. It can be deduced from the results that the difference between actual and perturbed shared data in default setting of chaincode and EDH-IIoT is decreasing with increasing the value of $\epsilon_{t}$. This is in line with the fact that increasing the differential privacy budget decreases the generated magnitude of calibrated noise and therefore, the difference between actual and perturbed data is low. More specifically, for $\epsilon_{t}$ = 3, the difference is very small as shown in Fig. \ref{dpriv3}. Therefore, the knowledge in the shared data is almost the same.  Similarly, if the distributor acts as an adversary and tries to link the information with his prior knowledge to get actual data then he will be confused because the shared data is perturbed.  Consequently, privacy is preserved at the same time. However, honest partners are impacted with noise added data. Furthermore, for smaller values of $\epsilon$ the difference is significant, and the accuracy is reduced as shown in Fig. \ref{dpriv1}. Therefore, it gives rise to a privacy and accuracy trade-off which is discussed in the next Section.  
\begin{figure}[htp]
\centering
\includegraphics[width=\linewidth]{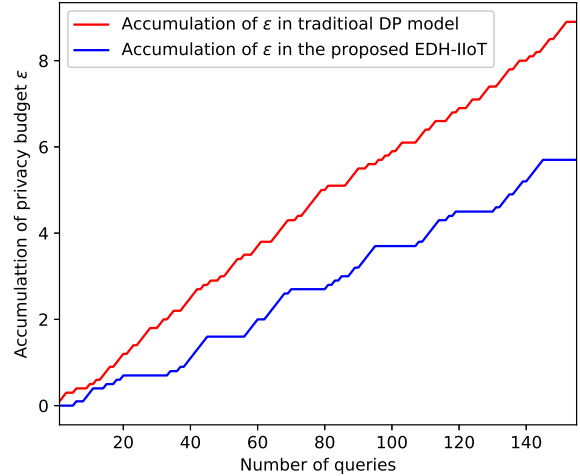}
\caption{Spending of differential privacy budget $\epsilon$ in traditional differential privacy model and EDH-IIoT. DP stands for differential privacy.}
\label{fig:bgt}
\end{figure}
\subsubsection{Relative Error}
\label{subsec:rerr}    
Noise addition into the query results reduces its accuracy in the proposed EDH-IIoT. The high the magnitude of noise the lower will be the accuracy and vice versa. To evaluate the accuracy of shared data, we evaluate the relative error $r$ in each query response. According to \cite{rerror}, the relative error $r$ can be defined as following:\\
\begin{align}
r = \frac{|a - a^{'}|}{a} \times 100\% && \text{(by \cite{rerror})}
\label{eq:er}
\end{align}
Here, $a$ is the actual data (query response) evaluated by the chaincode and $a^{'}$ is the noise added response of the query. Furthermore, to convert the relative error $r$ to percentage, we multiplied its value by 100.\\\\
The average relative error $r$ is the average of the individual relative errors of all queries. To evaluate relative error $r$, a total of 150 queries are sent to the SUT and the $\epsilon_{t}$ is varied from 1 to 5. The results obtained from this experiment are shown in Fig. \ref{fig:relerr}. The results show that the default setting of HF has $r$ = 0 because it shares the actual data. On the other hand, the noise added data in EDH-IIoT has $r > 0$.
Furthermore, the $r > 0$ reduces the accuracy of the shared data. Similarly, the accuracy of shared data increases as we go from $\epsilon_{t}$ = 1 to $\epsilon_{t}$ = 5. The minimum accuracy is 97.05\% whereas the maximum accuracy is 99.43\% for $\epsilon_{t}$ = 1 and $\epsilon_{t}$ = 5, respectively. The results also suggest that in order to get a balance between privacy and accuracy, the data sharing parties should agree on the selection of a suitable value of $\epsilon_{t}$. The privacy-accuracy trade-off is presented in Table \ref{tab:priv-acc}. 
\begin{table}[hbp]
\caption{Privacy and data accuracy trade-off}
\begin{center}
\begin{tabular}{|c|c|c|}
\hline
\textbf{Privacy budget $\epsilon_{t}$} & \textbf{Relative error $r$ (\%)} & \textbf{Accuracy (\%)}\\
\hline
1 & 3.85\% & 97.05\% \\
\hline
2 & 1.87\% & 98.55\% \\
\hline
3 & 1.29\% & 99.03\% \\
\hline
4 & 0.96\% & 99.27\% \\
\hline
5 & 0.77\% & 99.43\% \\
\hline
\end{tabular}
\label{tab:priv-acc}
\end{center}
\end{table}
\begin{figure*}[htp]
\centering
\begin{subfigure}[t]{0.48\linewidth}
\centering
\includegraphics[width=\linewidth]{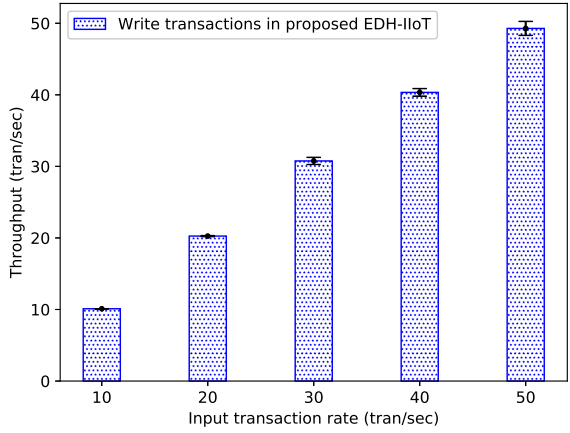}
\caption{}
\label{fig:thpputwrite}
\end{subfigure}
\begin{subfigure}[t]{0.48\linewidth}
\centering
\includegraphics[width=\linewidth]{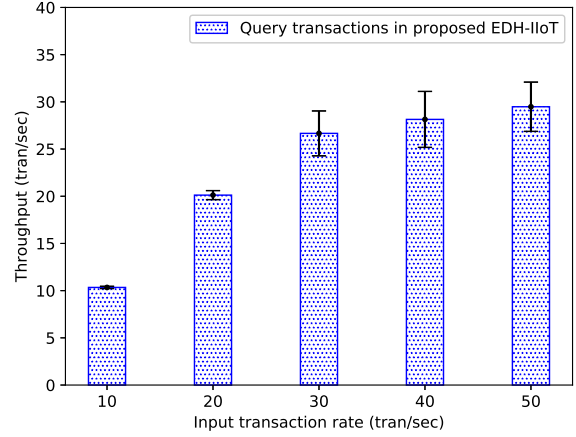}
\caption{}
\label{fig:thputquery}
\end{subfigure}
\caption{Evaluation of throughput in EDH-IIoT. The results are within 95\% of confidence interval.}
\label{fig:thput}
\end{figure*}
\begin{figure*}[htp]
\centering
\begin{subfigure}[t]{0.48\linewidth}
\centering
\includegraphics[width=\linewidth]{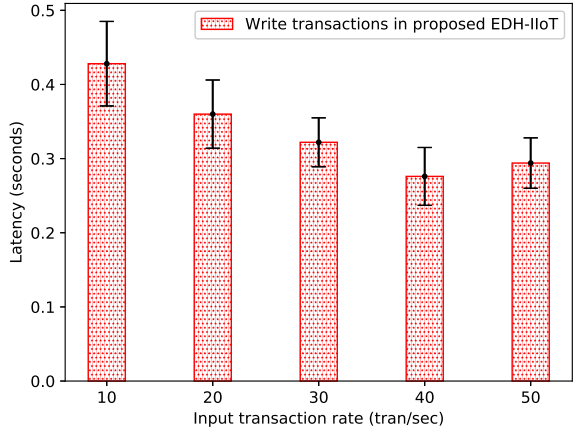}
\caption{}
\label{fig:latencywrite}
\end{subfigure}
\begin{subfigure}[t]{0.48\linewidth}
\centering
\includegraphics[width=\linewidth]{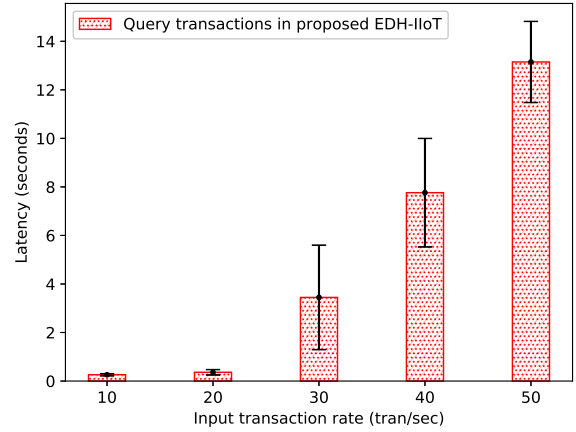}
\caption{}
\label{fig:latencyquery}
\end{subfigure}
\caption{Evaluation of latency of transactions in EDH-IIoT. The results are within 95\% of confidence interval.}
\label{fig:latency}
\end{figure*} 
\subsubsection{Tracking and Spending of Differential Privacy Budget $\epsilon$}
\label{subsub:trackandspend}
The tracking and spending of the differential privacy budget $\epsilon$ are evaluated by varying $\epsilon$ randomly in the range 0.01-0.12. Similarly, it is assumed that the data requester is known to the data provider because of the permissioned blockchain, i.e., every participant is known. In this experiment, a total of 155 query transactions were sent to the SUT and the spending of privacy budget is recorded for both traditional differential privacy model and EDH-IIoT. The results obtained are presented in Fig. \ref{fig:bgt}. The results show that the accumulation of privacy budget $\epsilon$ is 8.9 and 5.7 for traditional differential model and EDH-IIoT, respectively. The reason is that in traditional models, algorithm \ref{alg:algreuse} is missing which reuse the privacy budget for repeated queries. Consequently, the traditional model utilizes fresh privacy budget for each query which results in high value of accumulated privacy budget $\epsilon$. As a result, EDH-IIoT decreases the utilization of privacy budget $\epsilon$ by 35.96\%.\\
Furthermore, the proposed EDH-IIoT enables data provider to respond with different privacy preservation guarantees to the queries from various data requesters with different privacy requirements. In the same manner, it increases the number of potential query responses through efficient utilization of the privacy budget $\epsilon$. Furthermore, recording the spending and tracking of privacy budget $\epsilon$ ensures the associated data holders such as customers and employees that their privacy is not compromised. Consequently, the guarantee of intended privacy preservation is validated, i.e., the spending of privacy budget $\epsilon$ doesn't exceed the maximum limit $\epsilon_{t}$. 
\subsubsection{Throughput}
\label{subsec:thp}
We evaluate the throughput of EDH-IIoT by testing the SUT with varying input transaction rate between 10 and 50 tran/sec for two types of transactions which are write and query transactions. In this experiment, we sent 500 write and 755 query transactions to the SUT and varied the input transaction rate between 10 and 50 tran/sec. The results for both types of transactions are presented in Fig. \ref{fig:thput}. It is evident from the results that increase in input transaction rate causes an increase in the throughput for both types of transaction. The gentle increase in the throughput is the result of more input transactions to the SUT per unit time which results in more output transactions under the maximum transactions processing capacity of the SUT. It is clear from Fig. \ref{fig:thput}, that input transaction rate of 50 tran/sec results in a maximum throughput of almost 50 tran/sec for write transactions in EDH-IIoT. Similarly, in case of query transactions, the maximum throughput of 30 tran/sec is obtained for an input transactions rate of 50 tran/sec. Furthermore, the maximum throughput of query transactions is less than the maximum throughput of write transactions. The reason is that the query transactions are first evaluated on the ledger state and then the evaluation is recorded on the ledger at the same time.   
\subsubsection{Latency of Transaction} 
\label{subsec:lat}
The latency of transactions is evaluated by sending both write and query transactions to the SUT in the same manner as we did for throughput. In this experiment, a total of 500 write and 755 query transactions were sent to the SUT using Caliper tool. The results obtained are presented in Fig. \ref{fig:latency}. The results show that the latency for write transactions decreases up to 40 tran/sec of input transaction rate but goes up beyond that point. The fact is that until 40 tran/sec of input transaction rate, the SUT is under its maximum transactions processing capacity. Therefore, beyond this point the transaction latency shows increase for write transactions. Similarly, a sudden increase in the latency of query transactions is seen for input transaction rate beyond 20 tran/sec. The reason is that beyond this point the network is not under its maximum transactions processing capacity which results in high latency for query transactions. \\
It is evident from the evaluation of the proposed EDH-IIoT that it performs better than the default setting of chaincode of HF by enabling privacy preservation of sensitive data during the data sharing. By using a suitable value of $\epsilon_{t}$, the supply chain partners get the required knowledge from the query responses whereas exposure of business secrets and learning the lifestyle, spending habits, and financial status of the associated customers by the adversaries are avoided. Furthermore, the tracking of privacy budget $\epsilon_{t}$ enables the data sharing partners to get ensure that the privacy preservation is guaranteed as intended. The proposed EDH-IIoT achieves 97.05\% of accuracy in the shared results for $\epsilon_{t}$ = 1 which gives sufficient privacy guarantee for statistical COUNT queries. In this way, a data sharing environment between the supply chain partners in IIoT is established in a privacy preserving manner. The real-time data access from other partners provides key insights to other partners in the chain. Moreover, EDH-IIoT captures the real-world transactions in a blockchain-based supply chain network which consists of two organizations. Five customers were simulated to execute different purchase transactions. Through the simulation results, we validated that EDH-IIoT retains the high transaction processing rate of HF.         
\section{Conclusion} 
\label{sec:con}
In this work, a differentially private enhanced permissioned blockchain for private data sharing in the scenario of supply chain in IIoT is proposed which is titled as EDH-IIoT. First, a data sharing scenario in supply chain is presented in which the partners/organizations share and access data from the rest of the blockchain participants to incentivize their individual performance. The channels and private data collections along with querying mechanism of HF were used to enable data sharing. Subsequently, the problem of revealing sensitive information such as business secrets and personal information of customers was explored. Second, a solution was proposed by improving the query response mechanism of HF using the integration of differential privacy into the chaincode (smart contract). Moreover, the problem of privacy degrading through multiple queries on the same dataset (ledger data in this case) is also investigated. Furthermore, a tracking and spending mechanism of differential privacy budget was proposed to avoid the degrade of privacy preservation. Finally, the proposed work (EDH-IIoT) is evaluated extensively to verify the improvements in terms of privacy preservation by modeling linking and composition privacy attacks, efficient utilization of privacy budget and transactions processing. The results suggested that a privacy preservation guarantee of $\epsilon_{t}$ = 1 was achieved with 97.05\% of accuracy in the shared data. Similarly, the utilization of privacy budget $\epsilon$ was reduced by 35.96\% through reusing of privacy budget for repeated queries. Moreover, the proposed EDH-IIoT maintains the high transactions processing of HF.\\
In the future, we plan to classify the adversaries based on the threat level (high, medium, and low) and select the value of $\epsilon$ accordingly. Furthermore, we will use machine learning to dynamically select the most suitable value of $\epsilon$ which will help in minimizing the waste of total privacy budget $\epsilon_{t}$. Moreover, we plan to establish a dynamic trade-off between privacy preservation of sensitive data and the utility of data which will result in a sustainable privacy preserving supply chain model in the context of IIoT.   
\ifCLASSOPTIONcaptionsoff
  \newpage
\fi

\bibliographystyle{IEEEtran}
\bibliography{main.bbl}

\end{document}